\documentclass{article}

\usepackage[T1]{fontenc}
\usepackage[utf8]{inputenc}
\usepackage{lmodern}
\usepackage{microtype}

\usepackage{amsmath}
\usepackage{amssymb}
\usepackage{amsthm}
\usepackage{mathtools}
\usepackage{enumerate}
\usepackage[dvipsnames]{xcolor}

\usepackage{multirow}
\usepackage{empheq}
\usepackage{listings}
\usepackage{color}

\usepackage{graphicx} % Required for including pictures
\usepackage[font=small,labelfont=bf]{caption}

\usepackage[square,comma,numbers]{natbib}

\usepackage{hyperref}

\numberwithin{equation}{section}
\def\reff#1{{\rm(\ref{#1})}}

\def\be{\begin{equation}}
\def\ee{\end{equation}}

\def\bea{\begin{eqnarray}}
\def\eea{\end{eqnarray}}

\def\bea*{\begin{eqnarray*}}
\def\eea*{\end{eqnarray*}}

\def\cA{{\mathcal A}}

\def\cC{{\mathcal C}}

\def\cF{{\mathcal F}}

\def\cK{{\mathcal K}}

\def\cM{{\mathcal M}}
\def\cN{{\mathcal N}}

\def\cS{{\mathcal S}}

\def\cT{{\mathcal T}}

\def\cX{{\mathcal X}}

\def\cZ{{\mathcal Z}}

\def\E{\mathbb{E}}
\def\F{\mathbb{F}}

\def\R{\mathbb{R}}

\def\eps{\epsilon}

\def\reff#1{{\rm(\ref{#1})}}
\def\Om{\Omega}

\theoremstyle{plain}
\newtheorem{theorem}{Theorem}[section]

\newtheorem{proposition}[theorem]{Proposition}
\newtheorem{assumption}[theorem]{Assumption}

\newtheorem{example}[theorem]{Example}

\def\d{\mathrm{d}}

\newcommand{\ceq}{\mathrm{ce}}

\usepackage{fancyhdr}
\title{Deep Stochastic Optimization in Finance}
\author{A.\ Max Reppen\footnote{Questrom School of Business, Boston University, Boston, MA, 02215, USA, email: {\texttt{amreppen@bu.edu}}.}
\and H. Mete Soner\footnote{Department of Operations Research and Financial
Engineering, Princeton University, Princeton, NJ, 08540, USA, email: 
{\tt soner@princeton.edu}}
\and Valentin Tissot-Daguette\footnote{Department of Operations Research and Financial
Engineering, Princeton University, Princeton, NJ, 08540, USA, email: 
{\tt v.tissot-daguette@princeton.edu}}}

\date{\today}

\begin{document}

\maketitle
\vspace{5pt}

\abstract{This paper
outlines, and through stylized
examples
evaluates  a novel and highly effective computational technique
in quantitative finance.  
Empirical Risk Minimization (ERM) and neural networks
are key to this approach.  Powerful open source optimization
libraries allow for efficient implementations of this algorithm
making it viable in high-dimensional structures.
The free-boundary problems related to
American and Bermudan options showcase both 
the power and the potential difficulties that specific applications
may face.  The impact of the size of the 
training data is studied in a simplified Merton type problem.
The classical option hedging problem exemplifies
the need of market generators or large number
of simulations. }

\section{Introduction}
\label{s.introduction}

Readily available and  effective
optimization libraries such as Tensorflow or Pytorch
now make 
previously intractable 
regression type
of algorithms over hypothesis spaces with
large number of parameters 
computationally feasible. 
In the context of stochastic optimal control and
nonlinear parabolic partial differential
equations which have such representations, 
these exciting advances
allow for a highly efficient 
computational method.
This algorithm,
which we call \emph{deep empirical risk minimization}, 
proposed by 
E \& Han \cite{HE} and 
E, Jentzen \& Han \cite{HEJ},
uses artificial neural networks 
to approximate the feedback actions
which are then trained
by  empirical risk minimization. 
As stochastic optimal control
is the unifying umbrella for almost all hedging, portfolio or
risk management problems, and many 
models in financial economics,
this method is also highly relevant for 
quantitative finance.

Although
artificial neural networks as
approximate controls are widely used
in  optimal control and
reinforcement learning \cite{BT},
deep empirical risk minimization
simulates directly the system dynamics
and does not necessarily use dynamic programming.
It aims  to construct optimal actions and values offline
by using the assumed dynamics and the rewards structure,
and often uses market generators
to simulate large training data sets.
This key difference between reinforcement learning and 
the proposed algorithm ushers in essential
changes to their implementations and analysis as well.

Our goal
is to outline this demonstrably effective methodology,
assess its strengths and
potential shortcomings, and 
also showcase its power
through representative examples from  finance.
As verified in its numerous applications,
deep empirical risk minimization is algorithmically quite flexible and 
handles well a large class of high-dimensional models, 
even  non-Markovian ones, and 
adapts to complex structures with ease.
To further illustrate and evaluate these properties,
we also study three classical problems of finance
with this approach.
Additional examples from nonlinear partial differential equations and 
stochastic optimal control are given in
the recent survey articles of
Fecamp, Mikael \& Warin \cite{FMW} 
and  Germain, Pham \& Warin \cite{GPW}.
They also provide
an exhaustive literature review.

Our first class of examples is the 
American and Bermudan options.
The analysis of these instruments
offer many-faceted complex experiments through which
one appreciates the potentials and the challenges. 
In a  series of papers,  Becker {\emph{et al.}}, \cite{BCJ,ETH2}
bring forth a complete analysis with
computable theoretical upper bounds through its known
convex dual.  They also obtain  inspiring computational results
in high dimensional problems such as Bermudan max-call
options with $500$ underlyings.
Akin to deep empirical risk minimization is the 
seminal regression on Monte-Carlo
methods that were developed for the American options
by Longstaff \& Schwartz \cite{LS} and Tsitsiklis \& van Roy \cite{TvR}.
Many of their refinements, as delineated in the 
recent article of Ludkovski \cite{Lu},
make them not only  textbook topics but
also standard industrial tools.  Still, the 
deep empirical risk minimization approach to 
optimal stopping  has some advantages
over them,
including its
effortless ability to incorporate
market details and frictions,
and to operate in high-dimensions
as caused by state enlargements needed
for  path-dependent claims.
An example of the latter is the American options with rough volatility models
as studied by Chevalier \emph{at.~al.}~\cite{CPZ}.  They
require infinite-dimensional spaces
and their numerical analysis
is given in  Bayer \emph{et al.}~\cite{BTW}.
Other similar examples can be found in \cite{BCJ,ETH2}.

For  interpretability of our results,
we base the stopping decisions on a surface 
separating the `continuation' and `stopping' regions, and approximate 
directly this boundary -   
often called the \emph{free boundary} -
by an artificial neural network. 
Similarly for the same reason, Ciocan \& Mi\u{s}ic \cite{CM} 
compute the free boundary directly,
by using
tree based methods.
An additional benefit of this geometric approach to American
options is to construct a tool that
can also be effectively used for financial problems with discontinuous 
decisions such as regime-switching or transaction costs,
as well as non-financial applications.
Indeed,  the computation of the free-boundary is 
an interesting problem independent
of applications to finance.  
Recently, deep Galerkin method \cite{SS} is used
to compute the free boundary arising in the classical
Stephan problem of melting ice \cite{WP}.

Our numerical results, reported in the subsections
\ref{ss.one} and \ref{ss.maxcall} below,
show that 
natural problem 
specific modifications  
enable the general approach to yield excellent results 
comparable to the ones 
achieved in \cite{BCJ,ETH2}.
The free boundaries that we compute  for 
the two-dimensional max-call options
also compare to the results by
Broadie \& Detemple \cite{BD} and by Detemple \cite{Dbook}.
An important step in our approach
is to replace the 
stopping rule given by the 
sharp interface by a relaxed stopping rule 
given by a \emph{fuzzy boundary} as 
described in the subsection \ref{ss.fuzzy}.
Further analysis and the results of our free-boundary 
methodology are given in our future manuscript \cite{RST}.

Our second example of 
classical quadratic hedging \cite{S1} is undoubtedly
one of the
most compelling benchmark for any computational
technique in quantitative finance.  Thus, the evaluation
of the deep empirical risk minimization algorithm on this
problem,  imparts
valuable insights. 
Readily,  B\"uhler \emph{et al.}, \cite{BGTW,BGTWM}
use this approach for multidimensional Heston type
models, delivering convincing evidence for the flexibility and the scope
of the algorithm, particularly in high-dimensions.
Hur{\'e} \emph{et al.}, \cite{HPBL} and 
Bachouch \emph{et al.}, \cite{BHLP} also
obtain equally remarkable results for the 
stochastic optimal control using empirical minimization as
well as other hybrid algorithms partially based on 
dynamic programming.
Extensive numerical experimentations are also carried 
out by Fecamp \emph{et.~al}.~\cite{FMW} in an incomplete
market that models the electricity markets  
containing a non-tradable volume risk \cite{War}.
Ruf \& Wang \cite{RW} apply
this approach to market data of   S\&P 500 and Euro Stoxx 50 options.
In all these applications, variants of  the quadratic hedging error
is used as the  loss function.

To highlight the essential features,
we focus on a simple frictionless market with  Heston dynamics,
and consider  a vanilla Call option
with quadratic loss.
In this setting, we analyze both the pure hedging problem 
by fixing the price at a level lower than its 
known value and also the pricing and hedging problem by
training for the price as well.
By the well-known results of 
Schweizer \cite{S0,S1} and 
F\"ollmer \& Schweizer \cite{FoS}, we know that 
the minimizer of the analytical 
problem in the continuous time
is equal to 
the price obtained by Heston \cite{Hes}
as the discounted expected value under the risk neutral measure with
the chosen market risk of volatility risk.
Our numerical computations verify these results as well.

As the final example,
we report the results of
an accompanying paper of the first two authors \cite{RS}
for a stylized
Merton type problem.  With simulated data,
the numerical results once again 
showcase
 the flexibility and the scope
of the algorithm, in this problem as well.  We also observe that
in data-poor environments,
the artificial neural networks have an
amazing capability to over-learn the data
causing poor generalization.
This is one of the key results
of \cite{RS} which was also observed  in \cite{LPP}.  
Despite this potential,
as demonstrated by our experiments, continual
data simulation can overcome this difficulty 
swiftly.

In this paper, we only discuss the properties
of the algorithms that are variants 
of the deep empirical risk minimization.
The use of artificial neural networks or statistical machine
learning is of course not limited to this approach.
Indeed, starting from \cite{HLP}
and especially recently, 
artificial neural networks have been 
extensively employed in quantitative finance.
In particular, kernel methods
are applied to portfolio valuation in \cite{BF}, and to
the density estimation in \cite{FMS}.
Gonon \emph{et.~al}.~\cite{GMX} use the methodology
to study an equilibrium problem
in a market with frictions.
For further results and more information, 
we refer to the recent survey of Ruff \& Wang \cite{RWs}
and the references therein.

The paper is organized as follows.
The next section formulates  the control 
problem abstractly covering  many important
financial  applications. The description of the algorithm follows.
Section \ref{s.boundary} is about the American and Bermudan options.
The quadratic hedging problem
is the topic of Section \ref{s.options}.
Finally, the numerical examples related to the simple Merton
problem are discussed in Section \reff{s.merton}.
\vspace{10pt}

\noindent
{\bf{Acknowledgements.}} Research of the second and the third authors 
was partially supported by the National Science Foundation grant
 DMS 2106462.

\section{Abstract problem}
\label{s.problem}

Following the formulation of \cite{RS},
we start with  a $\cZ \subset \R^d$ valued stochastic process 
$Z$ on a
probability space
$\Om$.  This process drives the dynamics 
of the problem, and in all financial examples that we consider
it is the related to
the stock returns.  For that reason,
in the sequel, we refer to $Z$ as the
\emph{returns process}, although 
they may be logarithmic returns in some cases.  
Investment or hedging decisions are made
at $N$ uniformly spaced discrete time points labeled by 
$k=0,1,\ldots, N$ and let

\def\wht{\widehat{\cT}}
$$
\cT:= \{0,1,\ldots,N-1\},
\qquad
\wht := \{0,1,\ldots,N\}.
$$
We use  the notation $Z=(Z_1,\ldots,Z_N)$ and set $Z_0=0$.
We further let $\F=(\cF_t)_{t=0,\ldots,N}$ be the filtration
generated by the process $Z$.  The $\F$-adapted controlled state
process $X$ takes values in another Euclidean space $\cX$
and it may include all or some components of 
the uncontrolled returns process $Z$.

In the financial examples, the state
 includes the marked-to-market
value of the portfolio and  maybe other relevant quantities.
In a path-dependent structure, we would be forced to 
include not only the current value of the portfolio and 
the return, but also some past values as well (theoretically,
we need to keep all past values but in practice one 
stops at a finite point).  In illiquid markets,
the portfolio composition is also included into the state
and even the order-book might be 
considered.   We assume that the state
is appropriately chosen so that
\emph{the relevant decisions
are feedback functions of the state alone}
and we optimize over feedback decisions or controls.  
Thus, even if the original problem is low dimensional 
but non-Markov, one is forced to expand the state 
resulting in a high-dimensional problem.

We denote the 
set of possible actions or decisions by $\cA$.
While the main decision variable 
is the portfolio composition, several other quantities 
such as the speed of the change of the portfolio could be included. 
Then, a feedback decision
is a continuous function
$$
\pi : \cT \times \cX  \mapsto \cA.
$$  
We let $\cC$ be the set of all such functions.
Given $\pi \in \cC$, the time evolution
of the state vector is then completely described as a function
of the returns process $Z$.  Hence, 
all optimization problems that we consider have the 
following form,
$$
\text{minimize}\quad
v(\pi):= \E\left[\, \ell(\pi,Z)\, \right],
\qquad
\text{over all}\ \pi \in \cC,
$$
where $\ell$ is a nonlinear function.
We refer the reader to \cite{RS}
for a detailed derivation of the above formulation
and  several examples.  
Although the cost function $\ell$
could be quite complex to express analytically,
it can be easily evaluated by simply mimicking the 
dynamics of the financial market.
Hence, computationally they are 
straight-forward to compute and 
all details of the markets can be easily coded 
into it.

The goal is to compute the optimal feedback decision, $\pi^*$,
and  the optimal value
$v^*$,
$$
\pi^* \in  \text{argmin}_{\pi \in \cC} \, v(\pi),
\qquad
v^*:= \inf_{\pi \in \cC}\, v(\pi)\ = \ v(\pi^*).
$$
When the underlying dynamics 
is Markovian and the cost functional has
an additive structure, the above formulation
of optimization over feedback
controls  is equivalent to the standard
formulation which considers the larger class of
all adapted processes, sometimes called \emph{open
loop} controls \cite{FS}.  However, even 
without this equivalence, the minimization over the smaller
class of feedback controls
is a consistent and a well-defined problem,
and due to their tractability, feedback controls are widely used.
In this manuscript,
we implicitly assume that
the problem is well chosen
and the goal is to construct the 
best feedback control.

\section{The algorithm}
\label{s.derm}

In this section, we describe
the  \emph{deep
empirical minimization} algorithm proposed 
by \emph{Weinan E, Jiequn Han}, and \emph{Arnulf Jentzen} in \cite{HE, HEJ}.

A \emph{batch} 
$B:= \{Z^1,\ldots,Z^m\}$, with a size of $m$,  
is an i.i.d.~realization of
the returns process $Z$, where 
$Z^i=(Z^i_1,\ldots,Z^i_N)$ for
each $i$. 
We set
$$
L(\pi,B):= \frac1m\, \sum_{i=1}^m\, \ell(\pi,Z^i),
$$
and consider a set  of artificial neural networks
parametrized by,
$$
\cN=\left\{\, \Phi(\cdot;\theta) : \cT \times \cX 
\mapsto \cA\ \ :\
\theta \in \Theta\ \right\} \, \subset \, \cC.
$$
Instead of searching for a minimizer in $\cC$,
we look for a computable solution in the 
smaller set $\cN$.  That is, numerically we approximate the
following quantities: 
\begin{align*}
\theta^*&:= \theta^*_\cN \in \text{argmin}_{\theta \in \Theta}\, v(\Phi(\cdot;\theta)),
\\
v_{\cN} &:= \inf_{\theta \in \Theta}\, v(\Phi(\cdot;\theta))
= v(\Phi(\cdot;\theta^*)).
\end{align*}

The classical universal approximation result
for artificial neural networks \cite{Cybenko,Hornik}
imply, under some natural structural assumptions
on the function $\ell$, that $v_{\cN}$
approximates $v^*$ as the networks gets
larger as proved in \cite{RS}(Theorem 5).
This also implies that the performance of the trained feedback control
$\Phi(\cdot;\theta^*)$ is almost optimal.

The pseudocode of the algorithm to compute $\theta^*$ and $v^*$
is the following,
\begin{itemize}
\item \emph{Initialize}  $\theta \in \Theta$;
\item \emph{Optimize by stochastic gradient descent:}
for $n=0,1,\ldots$:
\begin{itemize}
\item Generate a batch $B:= \{Z^1,\ldots,Z^m\}$,
\item Compute the derivative $d:=\nabla_\theta\, L(\Phi(\cdot;\theta),B)$;
\item \emph{Update} $\theta \, \leftarrow \,
 \theta - \kappa  d$.
\end{itemize}
\item \emph{Stop} if $n$ is sufficiently large and the
improvement of the value is `small'.
\end{itemize}
In the above $\kappa$ is  the learning rate and
the stochastic gradient step
is done through an optimization library.  

The data generation
can be done through either an assumed
and calibrated  model, namely
a market generator, or
by random samples from a fixed
financial market data when sufficient 
and relevant historical data is available.
Although these two settings look
similar, one may get quite
different results in these two cases,
even when the fixed data set is large.
One of our goals is to better
understand this dichotomy between these two data regimes
and the size of the data needed for reliable results.
Theoretically,
when the simulation capability is not limited and data is continually
generated,
the above algorithm should
yield the  desired minimizer $\theta^*$ and the corresponding
optimal feedback decision $\Phi(\cdot,\theta^*)$.
However, with a fixed data set, 
the global minimum over $\cN$ is almost always
strictly less than $v^*$,
and the large enough networks will eventually
gravitate towards this undesirable
extreme point which would be over-learning
the data as already observed and demonstrated in \cite{RS}.

\section{Exercise boundary of American type options}
\label{s.boundary}

American and Bermudan options are particularly central to 
any computational study in quantitative finance as they 
pose difficult and deep challenges, and
they serve as an important benchmark 
for any new numerical approach.
Methods successful in this setting often
generalize to other problems as well.
Indeed, the seminal regression on Monte-Carlo
methods that were developed for the American options
by Longstaff \& Schwartz \cite{LS} and Tsitsiklis \& van Roy \cite{TvR}
have not only become industry standards in few years, 
but they have also shed insight
into other problems as well.  
Together with
rich improvements developed over the past decades,
they can now handle many  Markovian problem with ease.  However,
the key feature of
these algorithms is a projection onto
a linear subspace, and this space must grow
exponentially with the dimension of the ambient space, making
high-dimensional problems out of reach of this 
otherwise powerful technique.
Examples of such high-dimensional problems are
financial instruments on many underlyings 
modeled with many parameters,
path-dependent options, or non-Markovian models,
all  requiring
state enlargements and resulting in vast state spaces.

\subsection{Problem}
\label{ss.problam}

As well known the problem is 
to decide when to stop and collect 
the pay-off of a financial contract. Mathematically,
for $t \in \wht=\{0,\ldots,N\}$, let  $S_t \in \R_+^d$ be the stock 
value at the $t$-th trading date and
$\varphi : \R_+^d \mapsto \R$
be the pay-off function.  With a given
 interest  rate $r>0$, the problem is
 $$
 \text{maximize}\ \
 v(\tau):= \E\left[
 \, e^{- r\tau}\, \varphi(S_\tau)\, \right],
 $$
 over all $\wht$-valued stopping
 times $\tau$. We use the filtration generated by the 
 stock price process to define the stopping times.
 It is classical that the expectation is taken under the risk neutral 
 measure.  
 
We assume that $S$ is Markov
and the pay-off is a function of
the current stock value.  When it is
not,  then we need to enlarge
the state space.  In factor models like Heston or SABR, 
factor process is included.
In non-Markovian
models like the fractional Brownian motion, 
 past values the stock are added as in 
 \cite{BTW,BCJ,ETH2}.
In look-back type options, 
the minimum or the maximum of the 
stock process must be included in the state. 
We refer to \cite{RST} for the details
of these extensions.

We continue by defining
the price at all future points.
Recall that the filtration $\F$ is generated by
the stock price process. Let $\Xi_t$ be the set of all
$\F$-stopping times with values in $\{t,\ldots, N\}$.
At any $t \in \wht$,
 $s \in \R_+^d$, let $v(t,s)$
be the maximum value or the price 
of this option when  $S_t=s$, i.e.,
$$
v(t,s): = \max_{\tau \in \Xi_t}\ 
\E[ \, e^{-r(\tau-t)}\, \varphi(S_\tau)\ \mid \  S_t=s\, ].
$$
Then, $v(N,\cdot)=\varphi$
and  the 
the \emph{stopping region} is given by
\begin{equation}
\label{e.stop}
\cS := \left\{\, (t,s)\ :\
v(t,s)=\varphi(s) \, \right\}.
\end{equation}
Then the optimal stopping time 
is the first time to enter the region $\cS$, i.e.,
the following stopping time in $\Xi_t$
is a maximizer of the above problem:
$$
\tau^*:= \min \{\ u \in \{t,\ldots,N\}\ :\ (u,S_u) \in \cS\ \}.
$$
Notice that as $v(N,\cdot)=\varphi$, 
we always have
$(N,S_N) \in \cS$.  This implies that $\tau^*$ is well-defined
and is bounded by $N$.

Clearly, standard call or put options
are the main examples.
Many other examples that are also covered in 
the above abstract setting, including the max-call
option discussed below.
\vspace{5pt}

\begin{example}[Max-Call]
\label{ex.maxcall}

{\rm{ Let
$S_t =(S^{(1)}_t,\ldots,S^{(d)}_t)\in \R_+^d$ be a 
the stock process of $d$ many 
dividend bearing stocks.  We model it by a
$d$-dimensional
geometric Brownian motion 
with constant mean-return rate and a covariance matrix.
The pay-off of the max-call is given by,
$$
\varphi(S_t) = \left(\, \max_{i=1,\ldots,d}\, S^{(i)}_t\, - K\, \right)^+,
$$
where the strike $K$ is a given constant.
We study this example numerically in subsection \ref{ss.maxcall} below.
One can also consider max-call options with 
factor models with an extended state-space.}}
 
\end{example} 
\vspace{5pt}

\subsection{Relaxed stopping}
\label{ss.relax}

Quite recently, in a series of papers,  
Becker {\emph{et al.}}, \cite{BCJ,ETH2}
use deep empirical risk minimization in this context.
As the control variable is discrete (i.e., at any point in space, 
the decision is either `stop' or `go') and as the training or
optimization is done through a stochastic
gradient method, one has to relax the problem
before applying the general procedure.
We continue by first outlining this relaxation.
 
 In the relaxed version, we consider
 an adapted control process $p=(p_0,\ldots,p_N) $
 with values in $[0,1]$
 which is  the probability
 of stopping at that time conditioned on the event
 that the process has not stopped before $t$.  
 Because one has to stop at maturity, we have  $p_N=1$.
 Given the process $p$,  let $\xi_t^p$ be the 
 probability of stopping strictly before $t$.
Clearly,  $\xi_0=0$ and
 at other times it is defined recursively by,
 $$
 \xi^p_{t+1} = \xi^p_t+ p_t (1- \xi^p_t)\,
 =\, p_t + (1-p_t)\, \xi^p_t, \qquad t \in \cT.
 $$
 It is immediate that $\xi^p_t \in [0,1]$
and  is non-decreasing.
Also, if $p_t=1$, then $\xi^p_{s}=1$
for all $s >t$.   
The quantity $(1-\xi^p_t)$ is the unused
``stopping budget'',  and the relaxed stopping 
problem is defined by,
 \begin{equation}
 \label{e.vr}
 \text{maximize}\ \
 v_r(p):= \E\left[ \sum_{t=0}^N
 \, p_t\, (1-\xi^p_t)\, e^{r t}\, \varphi(S_t)\, \right],
 \end{equation}
 over all $[0,1]$-valued, adapted 
 processes $p$.
 The original problem of stopping is
 included in the relaxed one, as 
for any given stopping time $\tau$,
$p^\tau_t:= \chi_{\{t = \tau\}}$ yields
$\xi^\tau_t =  \chi_{\{t > \tau\}}$ and
consequently, $v(\tau)= v_r(p^\tau)$.
It is also known that this relaxation
does not change the value.

Becker {\emph{et al.}}, \cite{BCJ,ETH2} 
study the problem through this relaxation
and implement the deep empirical
risk minimization exactly
as described in the earlier section.
Additionally, 
using the known convex dual of the 
stopping problem, they are able
to obtain \emph{computable}
upper-bounds.  For many financial products of 
interest, they obtain remarkable results
in very high-dimensions.  They also
consider a fractional Brownian motion
model for the stock price.  As  for this example
there is no Markovian structure,
in their calculations the state is
all the past yielding an enormous state space.
Still the algorithm is tractable with 
computable guarantees.

\subsection{The free boundary}
\label{ss.boundary}

In most examples, the optimal stopping
rule is derived from a surface called the \emph{free boundary}.
For instance, the continuation region of a 
one-dimensional American Put option
is the epigraph of a function of time.
The stopping region 
of an American max-call option on the other hand,
is obtained by comparing the maximum of the stock
values to a scalar-valued function
as proved in Proposition \ref{p.geometry} below.
These stopping rules have the advantage of
being \emph{interpretable} \cite{CM} and easy
to implement.  Additionally,
free-boundary problems of this type
appear often in financial economics as
well as problems from other disciplines.
Thus numerical methods developed 
for the free-boundary of an American option could 
have implications elsewhere as well.

To be able to apply this method, we assume that
the stopping region $\cS$ has a certain 
structure.  Namely, we assume that
there exists two functions 
$$
\alpha : \R_+^d \mapsto \R,\qquad
\text{and}\qquad
F : \wht \times \R_+^d \mapsto \R,
$$
(recall that $ \wht=\{0,\ldots,N\}$) so that the stopping region of \eqref{e.stop} is given by,
$$
\cS= \{\, (t,s)\ :\ \alpha(s) \le F(t,s)\, \}.
$$  
More importantly, we also
assume that  $\alpha$ is given by the problem
and we only need to determine $F$ which we call the 
\emph{free boundary}. 
The following examples clarifies this assumption which holds in
a large class of problems.

\begin{example}
 {\rm{It is known that the stopping region
 of an American Put option with a Markovian 
 stock process  is given by
 $$
 \cS= \{ (t,s)\ :\ s\le f(t)\ \},
 $$
 for some function $f:[0,T] \mapsto \R_+$.  In this case, 
 $\alpha(s)=s$ and $F(t,s)=f(t)$.
 
 In the case of the max-call option,
 we show in Proposition \ref{p.geometry} below that
 for any $s=(s_1,\ldots,s_d)\in R_+^d$ with 
 $\alpha(s) =\max\{s_1,\ldots,s_d\}$, there exists a free boundary $F$.}}
 \qed
 
 \end{example} 
\vspace{5pt}
 
Given the above structure of the stopping 
region through the pair $(\alpha, F)$ the optimal 
stopping time is given by $\tau^*=\tau_F$,
where for any free boundary $F$,
$$
\tau_F\ := \ \min \ \{\ t \in \wht:\
\alpha(S_{t}) \le F(t,S_t)\ \}.
$$

 In this approach, the output of the artificial neural
 network is a scalar valued function $\Phi(\cdot;\theta)$
 of time and the state values, and it approximates
 the free boundary $F$.  Then for  any parameter
 $\theta$, the stopping time is
 $$ 
 \tau_\theta:= \tau_{\Phi(\cdot;\theta)}
 = \ \min  \{\ t \in \wht\ :\
\alpha(S_{t}) \le \Phi(t,S_t\, ;\, \theta)\ \}.
$$

 \subsection{Fuzzy boundary}
 \label{ss.fuzzy}
 
A sharp free-boundary
has the same problem of zero-gradients as 
the  original problem and its remedy is again a 
relaxation to allow for partial stopping.
Indeed, given a free-boundary $\Phi(\cdot;\theta)$
and a tuning-parameter $\eps>0$,
we define a fuzzy boundary region given by,
$$
F_{\Phi,\eps}:=\{\ (t,s)\ :\; -\eps \le \Phi(t,s ;\theta)-\alpha(s) \le \eps\, \}.
$$
If $\Phi - \alpha\le -\eps$ we stop, and if $\Phi - \alpha\ge \eps$ we continue,
and we do these
with probability one in each case.  But if the process falls
into the fuzzy region $F_{\Phi,\eps}$,  then as
in the relaxed problem, we assign a stopping probability
as a function of the normalized distance $d_t^\theta$ to the
sharp boundary $\{\Phi - \alpha=0\}$, i.e.,
$$
p^{\theta}_t := g(d_t^\theta),\qquad
\text{where}
\qquad
d_t^\theta= \frac{\Phi(t,S_t;\theta)- \alpha(S_t)}{\eps},
$$
and $g :[-1,1] \mapsto [0,1]$ is a fixed increasing, onto function.  
Linear or sigmoid-like functions are the obvious choices.
Once we compute the process
$p^{\theta}_t$,
the value 
corresponding to the parameter $\theta$ is $v_r(p^{\theta})$
with $v_r$ as in \eqref{e.vr}.
Hence, the relaxed free boundary problem is to train the network to
$$
\text{minimize}\ \ \theta\in \Theta \ \mapsto
 \ v_r(p^{\theta}).
 $$
The resulting trained artificial neural network
is  an approximation of the optimal
free boundary.

\subsection{American Put in one-dimension}
\label{ss.one}

As in \cite{BCJ,ETH2} we run the algorithm for an
 American put on a non-dividend paying stock whose price 
 process is modeled 
by a standard geometric Brownian motion with parameters
$$
S_0=K=40, T=1, \sigma =0.4, r =0.06,
$$
where as usual $S_0$ is the initial stock value,
$K$ is the strike, $\sigma$ is the volatility,
and $r$ is the risk-free rate.  
In this example, the state process is simply the stock process.

We are able to obtain accurate results
for the value as well as for the free boundary.
One typical result is given in Figure \ref{t1} below.
As the free boundary has a large derivative 
near maturity, we use a non-uniform mesh near
maturity.  Figure \ref{t1} uses 500 time points.
We also employ important sampling to ensure 
more crossings of the free boundary.
After the training is completed, the 
value corresponding to this trained free boundary
is computed by using the 
corresponding sharp interface.
Accurate price values are obtained rather easily.
 All of these
calculations are implemented by python in a personal laptop.

\vspace{10pt}

\begin{center}
\includegraphics[width=\linewidth,height=1.85in]{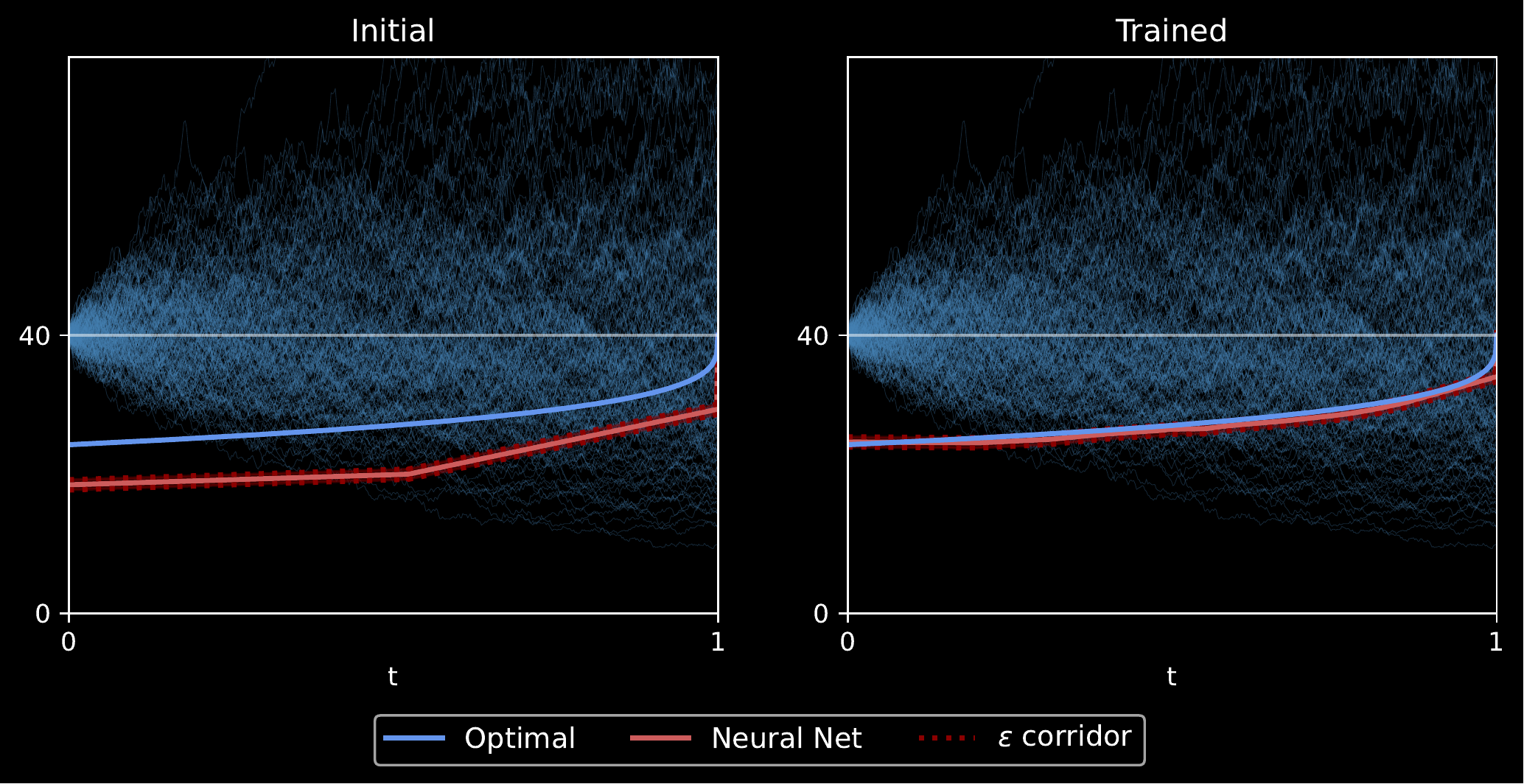}
%\captionof{figure}{Position}
\label{t1}
\captionof{figure}{Left figure is a random initialization
and the right one is the final trained boundary.  Dashed line
is the optimal calculated through a finite-difference scheme.
The price is $5.311$. }
\end{center}

\vspace{10pt}

\subsection{Max-Call Options}
\label{ss.maxcall}

In this subsection,
we consider the max-call
option studied in the seminal paper by
Broadie \& Detemple \cite{BD}
and also in the book by Detemple  \cite{Dbook}.
Let  $S_t \in \R^d$ be the  price process
of a dividend bearing stock.
The pay-off the max-call option at time $\tau$ is 
$$
\varphi(S_\tau)= \left(\, m(S_\tau) \, -\, K\, \right)^+ ,\qquad
$$
where the function $m:\R_+^d \mapsto \R_+$ is given by,
$$
m(s):= \max_{i=1,\ldots,d}\, \, s_i, \qquad s =(s_1,\ldots,s_d) \in \R_+^d.
$$

The main 
structural assumption needed 
is the natural sub-linear
dependence of the stock
prices on their initial values.

\begin{assumption}[Sublinearity]
\label{a.sublinear}
For $t \in \cT$,  $s \in \R_+^{d} $,
non-decreasing function $\phi: \R_+^d \mapsto \R$,
$\lambda \ge 1$ and a stopping time $\tau \ge t$,
$$
\E\left[ \phi(S_\tau) \mid S_t= \lambda s\right]\,
\le \,  \E\left[ \phi(\lambda\, S_\tau) \mid S_t= s\right].
$$
\end{assumption}

Above assumption is satisfied in all examples.
In fact, in most models the dependency 
on the initial data is linear.
Although in our numerical calculations,
we use a geometric Brownian motion model 
for the stock price process, the method also
applies more generally to all
factor models.  

We use this assumption to show that the 
stopping region has a certain geometric structure
which we exploit.  The following result
is already proved in \cite{BD}
and more generaly in \cite{RST}.
We provide its proof for completeness.
Let $\cS$ be as in \eqref{e.stop} and set
$$
\cK:= \left\{\, s \in \R_+^d\ :\
m(s)=1\, \right\}.
$$
Note that for any $s \in \R_+^d$, $\frac{s}{m(s)} \in \cK$.

\begin{proposition}
\label{p.geometry}
Consider the max-call option
in a market satisfying
 the Assumption \ref{a.sublinear}.  
Then, if $(t,s) \in \cS$,
then $(t,\lambda s) \in \cS$
for any $\lambda \ge 1$.
In particular, 
$$
\cS= \{\ (t,s) \ :\
m(s) \ge F(t, s/m(s))\ \},
$$
where  $F :\wht \times \cK \mapsto \R_+$
is given by,
$$
F(t,k):= \inf \left\{\, \rho >0\ :\ (t,\rho k) \in \cS\ \right\},
\qquad m \in \cM.
$$
\end{proposition}

Above result can be equivalently stated
 as the $t$-section
 $S_t:= \{\ s\in \R_+^d\ :\ (t,s) \in \cS\ \}$
 of the continuation region 
 being \emph{star-shaped} for every $t$.
 
 \def\sp{s^\prime}
 \begin{proof} 
  Suppose that $(t,s) \in \cS$ and $\lambda \ge 1$.
  As $\{(N,s)\ :\ s\in \R^d_+\} \subset \cS$,
if $t=N$, clearly $(t,\lambda s) =(N,\lambda s) \in \cS$.
So we assume that $t<N$.  Then, a
point $(t,\sp)$ is in $\cS$ if and only if
  $\sp >K$ and
 the following inequality is satisfied  for every $\tau \in \Xi_t$:
 $$
\E\left[\, e^{-r(\tau -t)}\, (S_\tau -K)^+\, |\, S_t=\sp\, \right]
\le  \sp-K.
 $$
 By Assumption \ref{a.sublinear},
\begin{align*}
\E\left[\, e^{-r(\tau -t)}\, (S_\tau -K)^+\, |\, S_t=\lambda s\, \right]
&\le
\E\left[\, e^{-r(\tau -t)}\, (\lambda S_\tau -K)^+\, |\, S_t=s\, \right]\\
&=
\E\left[\, e^{-r(\tau -t)}\, (\lambda [S_\tau -K]+(\lambda -1)K )^+\, |\, S_t=s\, \right]\\
 &\le
\lambda \, \E\left[\, e^{-r(\tau -t)}\, (S_\tau -K )^+\, |\, S_t=s\, \right]
+(\lambda -1)K\\
 &\le
 \lambda(s-K) +(\lambda -1)K\\
 &=(\lambda s -K).
 \end{align*}
 Hence, we conclude that $(t, \lambda s) \in \cS$. 
 \end{proof}

\subsubsection{Numerical Experiments}

We consider a max-Call option 
and in a geometric Brownian motion model under the risk
neutral measure,
$$
S_t \ =\ S_0\, \exp\left( (r- div) t + \sigma W_t - \frac12 \sigma^2t \right),
$$
with parameters
$$
K=100,\ S_0=90,100,110,\ \sigma =0.2,\ r =0.05,\ div =0.1,
$$
where the notation is as in the previous subsection and $div$ is the dividend rate.
We take the maturity to be $3$ years and $N=9$.  Thus, each
time interval corresponds to four months.  All these parameters
are taken from  \cite{BCJ,ETH2} to allow for comparison.
We also make qualitative comparison
to the results of \cite{BD}.

\begin{table}[ht]
  \centering
  \begin{tabular}{|c| c| c| c| c| c| c|}
 \hline
   {\bf{Runs}}&  {\bf{1}} &  {\bf{2}}&  {\bf{3}}&  {\bf{4}}&  {\bf{5}}&  {\bf{6}}\\
  \hline \hline
 {\bf{Price}} & 8.0747 & 8.0757 & 8.0710 & 8.0684 & 8.0670 & 8.0731 \\
 \hline
  {\bf{Stdev}}& 0.00305 & 0.00315& 0.00310& 0.00310& 0.00309& 0.00311 \\
 \hline
  \hline
   {\bf{Runs}} &  {\bf{7}}&  {\bf{8}}&  {\bf{9}}&  {\bf{10}} &  {\bf{Mean}} &  {\bf{Stdev}}\\
  \hline \hline
  {\bf{Price}} & 8.0686 & 8.0707 & 8.0620 & 8.0679 & 8.0699& 0.0040\\
 \hline
  {\bf{Stdev}}& 0.00306& 0.00308& 0.00307& 0.00311& - & - \\
 \hline
\end{tabular}
    \caption{\label{t.1}Ten experiments with $S_0=90$,
  batch size $2^{13}$, $7000$ iterations.
  Prices are calculated with $2^{23}$ simulations.
  Stdev in the third and sixth rows refer to the standard deviations of the Monte-Carlo
  simulations, while Stdev at the end is the standard deviation of the 
 calculated ten prices. }
  \end{table}

Table \ref{t.1} shows the results
with $d=2$, $S_0=90$,
batch size of $B=2^{13}$ and
$7,000$ iterations.
The corresponding price is computed after the training
is completed with $2^{23}$ Monte-Carlo simulations
using the sharp boundary instead of the fuzzy one.
Important sampling is used with a $1.4\%$ downward 
drift.
We repeated the experiment ten times
in a personal computer.
All of the results are within the $95\%$ confidence interval
$[8.053\, ,\, 8.082]$ computed 
in \cite{AB}.  The standard deviation
of each price computation is quite low.
Hence, the maximum of the values
is a lower bound for the price.

We also repeated the experiments of  \cite{BCJ,ETH2} 
in space dimensions $d=5, 10, 100$ with the above
parameters.  For each parameter set, we computed ten
prices exactly as described above.  The results reported in Table \ref{t.4} below
are in agreement with the results of  \cite{ETH2} (Table 9).
We should also note when $d$ is large, the maximum  of  many stocks 
have a very strong upward drift making the standard deviation 
of the rewards quite high.  

\begin{table}[ht]
  \centering
  \begin{tabular}{|c| c| c| c| c| c| c|}
 \hline
   {\bf{Dim.}}&  {\bf{$S_0$}} &  {\bf{Price}}&  {\bf{Std}}&  {\bf{Price in \cite{ETH2}}}&
    {\bf{Max Price}} & {\bf{Its Std}}\\
  \hline \hline
2 &  90 & 8.0699 &  0.0031 &8.068 &  8.0757 &  0.0040 \\
 \hline
 2 & 100  & 13.9086 &  0.0059 &13.901 & 13.9128&  0.0033 \\
 \hline
  2 & 110  & 21.3434 &  0.0059 &21.341& 21.3541 &  0.0104 \\
 \hline
  5 & 90  & 16.6187 &  0.0040 & 16.631 & 16.6238 &  0.0045 \\
 \hline
  5 & 100  & 26.1194 &  0.0259 & 26.147 &26.1644 &  0.0057 \\
 \hline
  5 & 110  & 36.7176&  0.0078 &36.774 & 36.7408 &  0.0078 \\
 \hline
  10 & 90  & 26.2130 &  0.0182 & 26.196 & 26.2362 &  0.0069 \\
 \hline
  10 & 100  & 38.2735 &  0.0538 & 38.272 & 38.3351 &  0.0089  \\
 \hline
  10 & 110  & 50.8350 &  0.0397 & 50.812 &50.8685 &  0.0081  \\
 \hline
  100 & 90  & 66.2460 &  0.4946 & 66.359&  66.6163&  0.0223 \\
 \hline
  100 & 100  & 82.5475 &  0.6463 &83.390  & 83.6563 &  0.0272  \\
 \hline
  100 & 110  & 98.9868&  0.0366 & 100.421 & 99.0575&  0.0353  \\
 \hline
\end{tabular}
 \caption{\label{t.4}Each price is the mean of ten
 experiments with parameters as in Table \ref{t.1}. 
 Max price is the maximum of ten
 experiments 
 with a standard deviation of the price calculation
 with $2^{23}$ Monte-Carlo simulations.}
  \end{table}

  The above table reports the average values
  for ten runs to be able to asses the possible variations.
  However, the maximum value among these
  ten runs is in fact a lower bound the actual price.
 As we computed these values with $2^{23}$ (roughly eight million)
 simulations, the standard devision of this price value is small.
 
In two dimensions, the stopping 
region can be visualized effectively.
Figures \ref{t.2}, \ref{t.3} are stopping regions
in two space dimensions
obtained with 
initial data $S_0=90$ and $S_0=100$.
Clearly the free boundary is independent of
the initial condition
and the below numerical results
verify it.  Also they are similar to those obtained in \cite{BD}.

\begin{center}
\begin{figure}[h]
\includegraphics[height=2in,width=5in]{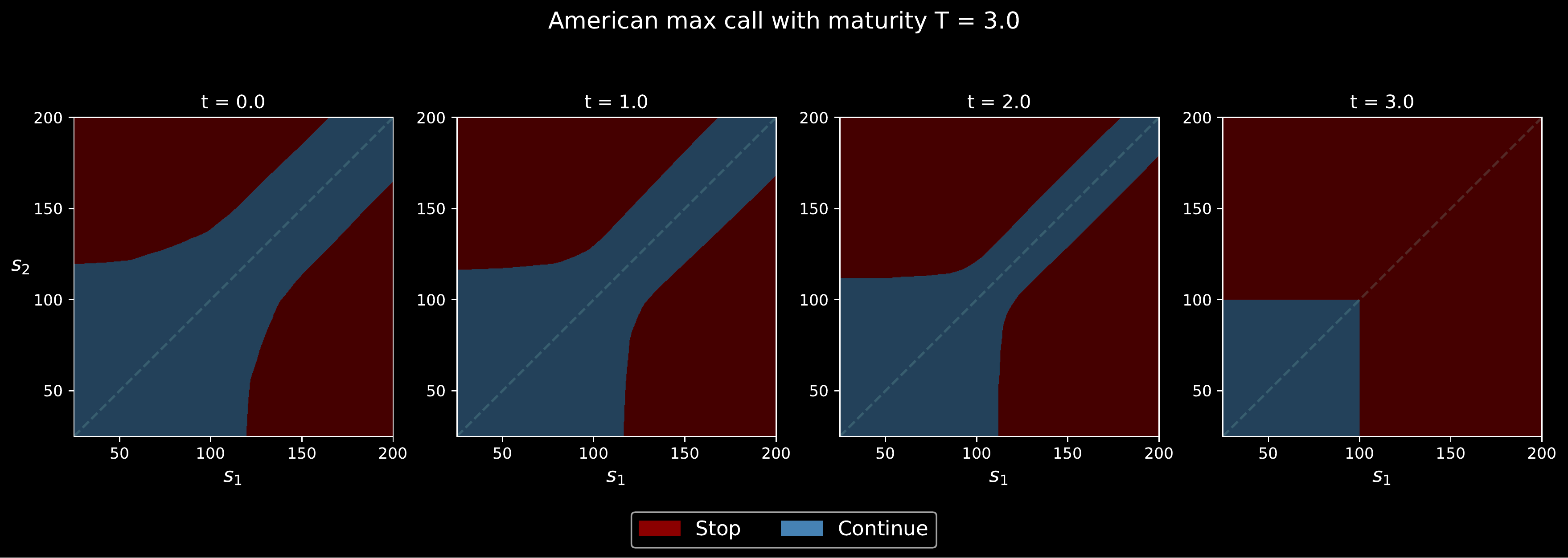}
\caption{\label{t.2}Evolution of the Free Boundary with $S_0=90$}
\end{figure}

\begin{figure}[h]
\includegraphics[height=2in,width=5in]{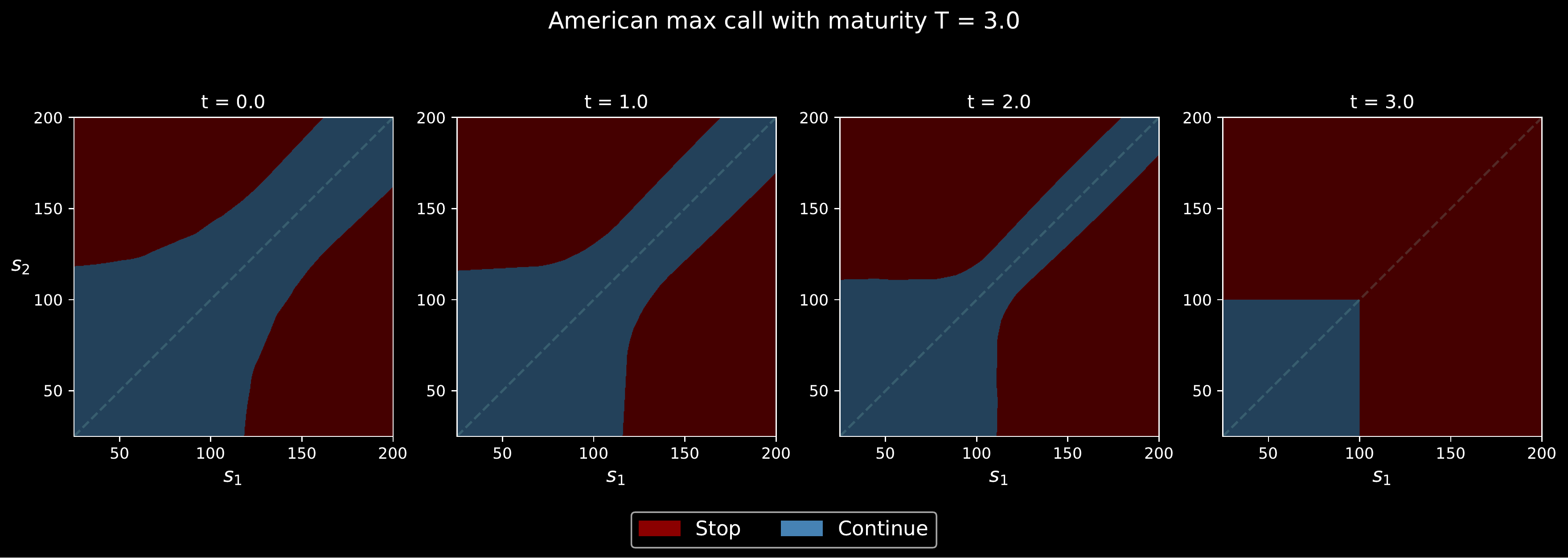}
\caption{\label{t.3}Evolution of the Free Boundary with $S_0=100$}
\end{figure}
\end{center}

\section{Valuation and Hedging}
\label{s.options}

We consider a European option
with stock process $S$ and pay-off $\varphi(S_T)$.
We consider the  Heston dynamics,
\begin{align*}
\d S_t & =S_t\, ( \mu \d t +\sqrt{v_t\, }\, \d W_t),\\
\d v_t &= \left( \kappa (\theta -v_t) - \lambda\, v_t \right)\, \d t + \sigma\, v_t \, \d \tilde{W}_t,
\end{align*}
where $W, \tilde{W}$ are one-dimensional Brownian motions with 
constant correlation of $\rho$, and the five Heston parameters
 $(\mu, \kappa, \theta, \sigma, \rho)$ 
 are chosen satisfying the Feller condition.
 In particular, we choose the market price of volatility risk parameter $\lambda$.

Let $p^*$ be the 
price of this claim,  and
$Z$ be the return process, i.e.,
\begin{equation}
\label{e.1}
Z_{t+1}: =\frac{S_{t+1}-S_{t}}{S_{t}},\qquad
t \in \cT.
\end{equation}
Further, let the feedback actions be the 
continuous functions
$$
\pi : \cT \times \R_+ \times \R \ \mapsto \ \R,
$$
representing the dollar amount invested in the stock.
The corresponding wealth process is given by,
\begin{equation}
\label{e.2}
X^{\pi,x}_{t+1} = (1+r)\, X^{\pi,x}_{t}  + \pi(t,X^{\pi,x}_{t},Z_{t})\,\left( Z_{t+1} -  r\, \right),
\quad t \in \cT,
\end{equation}
with initial data $X^{\pi,x}_0=x$. 

We first fix an initial wealth of $x <p^*$
and consider the following
\emph{pure-hedging problem}
of minimizing the square hedging error, i.e.,
\begin{equation}
\label{e.pure}
v^*(x):= \min_{\pi \in \cC}\, 
v(x,\pi),\qquad
\text{where}\qquad
v(x,\pi) := \E\left[ (\varphi(S_T) - X^{\pi,x}_T)^2\ \right].
\end{equation}

In the second problem, we minimize over $x$ as well, i.e,
\begin{equation}
\label{e.price}
v^*:= \min_{x\in \R}\, v^*(x) =  \min_{(x, \pi) \in \R \times \cC }\, 
v(x,\pi).
\end{equation}
As proved by F\"ollmer \& Schweizer \cite{FoS},
it is well-known that in continuous time
the solution to the second problem, $v^*$,  is equal
to the Heston price.  Thus,
for sufficiently fine discretization $v^*$
is close to  zero, 
$x^*$ is  close to the  known 
continuous-time Heston price.  Also
the numerical hedge $\pi^*$ must
be equal to the continuous time hedge.

If $r=0$, then, $X^{\pi,x}_t=x+X^{\pi,0}_t$ and the 
initial wealth $x$ only influences the mean of the hedging error.
Therefore, we expect that after an initial
adjustment to minimize the mean, the networks
would minimize the variance which is 
independent of the initial wealth.
This approximate reasoning indicates
that after an initial transient region,
both minimization problems may behave similarly when 
there is large data.

\subsection{Numerical results}
\label{ss.num1}

We implemented the above hedging problem in Julia's Flux \cite{innes2018flux} by parameterizing the portfolio at each time point, including the initial wealth level.
In particular, we hedge a call option with strike $K$, i.e., $\varphi(x) = (x - K)_+ = \max\{x - K, 0\}$.
Our implementation follows the scheme in Section~\ref{s.derm}, which we here describe in greater detail for this particular problem.

We see in \eqref{e.price} that the two quantities we optimize over are $x$ and $\pi$.
As $x$ is a scalar, we directly parameterize it with a 1-element tensor, which after optimization is the option price.
The policy $\pi$, however, can be approximated in various ways.
We here opt for a very direct method in which we represent it by a single neural network with time and stock data as inputs.
This contrasts \cite{BGTW}, where the authors discretize time and design one neural network per time point.
As we shall see, our implementation of a single neural network also performs well, with the additional benefit of allowing changes to the time discretization during training.
There are also other training differences between the two parameterizations, as, for instance, the one used here accomplishes a large degree of parameter sharing.
Nevertheless, a thorough account of these differences is outside the scope of the present paper.

Another detail of our implementation is that we write $\pi$ as a function of $t$ and $S_t$ instead of the formulation in \eqref{e.2}.
It is clear that the two are mathematically equivalent, although they could differ in training performance.
Ours is a naïve choice and we make it because we find it more natural, not because it necessarily leads to better performance.
The neural network is designed with two hidden layers of width 20 and with ReLU activation.
In-between layers, batch normalization is employed.%
\footnote{Although we believe that the following parameters are not crucial for replicating our results (because they were not tuned), we list them here for completeness:
batch size: 512;
optimizer: Adam with the Flux default parameters $(\eta, \beta_1, \beta_2) = (0.001, 0.9, 0.999)$;
and the number of epochs was a fixed value for which the training error of a typical run had plateaued.
%Although these parameters can affect the speed of convergence and the resulting error size, they should not affect whether convergence happens.
}

The results of our computations are presented in Table~\ref{table:heston_hedging}.
We compare our numerical solution to the Heston prices from \url{https://www.quantlib.org/}.
No significant tuning has gone into producing our values, and it is nevertheless clear that accurate prices are consistently attained.
We see, for instance, that the absolute error is approximately the same for all three strikes, which we argue is a consequence of
(i) not tuning the training parameters to each individual problem and
(ii) our hedging is in discrete time, which introduces a time discretization error.
Although this is only a one-dimensional problem, it gives credence to the method's effectiveness, effectiveness that does translate into higher-dimensional performance, as we illustrated for the American options problems.

\begin{table}[ht]
\begin{center}
\begin{tabular}{c c c c c}
$K$ & QuantLib price & Price & Avg.\ abs.\ error & Error std.\ dev. \\
\hline
90 & 10.076508 & 10.078163 & 0.001869 & 0.001174 \\
100 & 2.295405 & 2.295211 & 0.002018& 0.001065 \\
110 & 0.128136 & 0.127069 & 0.001971 & 0.001793
\end{tabular}
\caption{\label{table:heston_hedging} Hedging performance of a call option with strike $K$ in a Heston model with parameters $S_0 = 100$, $v_0 = 0.04$, $\kappa = 0.9$, $\theta = 0.04$, $r = \lambda = 0$ and $\sigma = 0.2$.
  The maturity $T = 1/12$ and the time interval is discretized in 22 steps.
  Each row lists the deep hedging price average over 100 runs along with the standard deviation over the same 100 runs.}
\end{center}
\end{table}

 Theoretically,
in continuous-time,
the optimal hedge is independent of the 
initial wealth.  We also studied this
by fixing the initial portfolio value
to the price and also to its half value.
One simulation of the trained hedge is given
in the Figure \ref{t.3} shows that
the dependence is minimal.

\begin{figure}[h]
\begin{center}
\includegraphics[height=2in,width=4in]{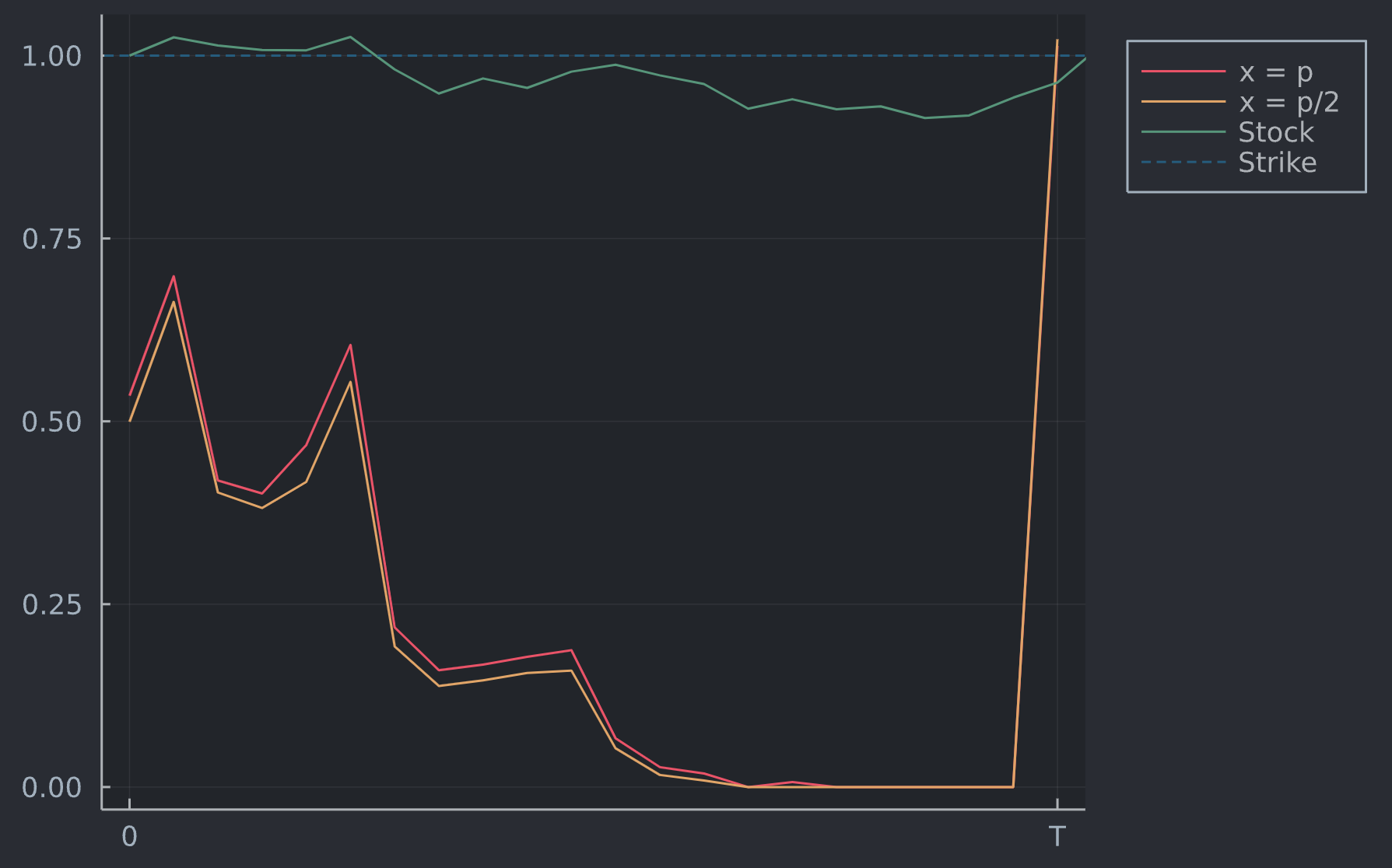}
\caption{\label{t.3}Optimal Hedges for the Heston model. 
The orange curve is the trained feedback hedge
with an initial wealth half
the option price,
while the red curve is trained
with initial wealth equal to the option price.
The stock price is rescaled to start at $S_0=1$. }
\end{center}
\end{figure}

\section{Merton Problem and Overlearning}
\label{s.merton}

In this section, we summarize the results of \cite{RS} by
the first two authors.
As in that paper,  to emphasize the essential features
of the algorithm,
a simple financial market without any frictions
and constant coefficients is considered. Additionally, consumption
is not taken into account.  All these details can  incorporated
into the model and problems
with complex market structures have already been studied
extensively by Buehler \emph{et.al.}~\cite{BGTW,BGTWM}.

Consider a stock price process  $S_t \in \R_+^d$ in discrete time
and assume a constant interest rate of $r$.  Let
the return process $Z$ be as in \eqref{e.1}
and $X^\pi=X^{\pi,x}$ be as in \eqref{e.2}. We suppress 
the dependence of the initial wealth $x$ for simplicity. Then, the classical 
investment  problem 
is to maximize
$v(\pi):= \E[U(X^\pi_T)]$
with a given utility function $U$.

In \cite{RS} it is proved that
the deep empirical risk minimization algorithm converges 
as the size of the training data gets larger.  On the other hand,
it is also shown that for fixed 
training data sets, larger and deeper neural networks have 
the capability of overlearning the data, however large it might be.
In such situations, the trained neural networks
while substantially over-perform the theoretical optimum on the training set,
they do not generalize and perform poorly on other data sets.

These theoretical results are demonstrated
in the following stylized example
with an explicit solution in \cite{RS}(Section 8). 
In that example, the utility function is taken to be the exponential
with parameter one,
and as the decisions are independent of the initial value for
these class of utilities, 
the initial value is fixed as one dollar.
To simplify even further, for one 
period this amount is invested uniformly on all stocks.
Then,  with
${\bf{1}}:=(1,\ldots,1)$,
$\pi_0= {\bf{1}}/d$ and $X_1= (Z_1 \cdot {\bf{1}})/d -r$ are uncontrolled,
and the  investment problem is to choose the feedback
portfolio $\pi_1(Z_1) \in \R^d$ 
so as to maximize
$$
v(\pi)= \E\big[ 1- \exp(- X^\pi_2)\big],
$$
where
$X^\pi_2= (1+r)X_1 + a(Z_1) \cdot (Z_2-r {\bf{1}})$.
The \emph{certainty equivalent} of a utility value $v<1$ given by
$$
\ceq(v):= \ln(1-v) \quad
\Longleftrightarrow
\quad
v= U(\ceq(v))
$$
is a more standard way of
comparing different utility values.
Indeed, agents
with expected utility preferences
would be indifferent
between an action $\pi$ and a cash amount of $\ceq(v(\pi))$
because the 
utilities of both positions are equal to each other.  Thus, for these agents
the cash equivalent of the action $\pi$ is $\ceq(v(\pi))$.

The following table \cite{RS}(Table 1) clearly
demonstrates overlearning.  In this experiments
 the training data of size $N=100,000$  and an artificial
 neural network with three hidden 
layers of width 10 is trained on this set for four or five               
epochs.  For each 
dimension the algorithm is run thirty times 
and 
Table \ref{tab:learning}  below reports 
the mean and the standard. deviation.
Although conservative stopping rules are employed 
in \cite{RS}, there is substantial overperformance increasing with dimension.

\begin{table}[ht]
  \centering
  \begin{tabular}{c|rr|rr}                
    \multirow{2}{*}{dims} & \multicolumn{2}{c|}{$p_{in}$ (\%)} 
    & \multicolumn{2}{c}{$p_{in} - p_{out}$ (\%)} \\
& $\mu\ \quad$ & $\sigma\ \quad$ & $\mu\ \quad$ & $\sigma\ \quad$ \\
    \hline
    100 & 10.12820 & 1.09290 & 23.67080 & 2.01177 \\
    85 & 8.38061 & 1.35575 & 20.16440 & 2.30489 \\
    70 & 7.32720 & 0.86458 & 15.62060 & 1.94043 \\
    55 & 5.05783 & 0.81518 & 10.93950 & 1.54431 \\
    40 & 3.74648 & 0.62588 & 7.91105 & 1.32581 \\
    25 & 2.11501 & 0.43845 & 4.58954 & 0.88461 \\
    10 & 0.53982 & 0.34432 & 1.46138 & 0.39078 \\
    \end{tabular}
  \caption{\label{tab:learning}Average  
  relative in-sample performance, and its comparison 
  to the out-of-sample performance
  with the above described conservative stopping rule.  
 Everything is in \% with training size of $N=100,000$ and
 three hidden layers of width 10.  
 The $\mu$ value
is the average of 30 runs
 and $\sigma$ is the standard deviation.}
\end{table}

\section{Conclusion}\label{sec13}

The deep empirical risk minimization
proposed by E, Han \& Jentzen \cite{HE,HEJ} 
provides a flexible and a highly effective tool
for stochastic optimization problems arising in
computational finance. 
Recent development of
optimization libraries make this algorithm 
tractable in very high dimensions allowing
to include important market details such as
factors and frictions, as well as models
with long memory.  Once a large 
training set is given, the algorithm 
mimics the market dynamics with
all its details.  This simple description 
together with powerful new computational
tools are keys to the power of the algorithm.
We have demonstrated above properties in three different
classes of problems.  As it is always the case,
each requires problem specific but natural modifications.
Moreover, the output can be designed to be 
exactly the decision rule that is under investigation.

The method on the other hand needs large 
data sets for reliable results.  In the financial 
setting this essentially limits its scope to
model driven markets with an unlimited simulation capability.
However, due to its seamless transition to more
complex structures, more interesting parametric models
are now feasible.
Thus, on-going research on market generators 
will be an important factor on further developments.

\bibliographystyle{abbrvnat}
\bibliography{ANN-ERM}
\end{document}